\documentclass{PoS}

\title{Energy Dependance of~Azimuthal Correlations (and~more)
	in~Nucleus--Nucleus Collisions at~the~CERN SPS}

\ShortTitle{Energy dependence of azimuthal correlations}

\author{\speaker{Marek Szuba} for the NA49 Collaboration\\
        Faculty of Physics, Warsaw University of Technology,
        Koszykowa 75, 00-662 Warszawa, Poland\\
        E-mail: \email{Marek.Szuba@if.pw.edu.pl}}

\abstract{
Analysing two-particle azimuthal and pseudorapidity--azimuth
correlations of high-$p_T$ hadrons, experiments at the RHIC have
obtained results --- ``the double hump'', ``the ridge'' etc. --- which
provide interesting information about properties of the hot, dense medium
produced in central high-energy heavy-ion collisions as well as of its
interaction with products of hard scattering. However, not all of the
observed effects have been fully explained by theory or models yet; such
explanations are hampered in part by the lack of good reference from lower
collision energies. In an attempt to further our understanding of angular
dihadron correlations NA49 has, as one of the first experiments at the SPS,
performed a number of scans of these correlations in \textit{Pb+Pb}, \textit{Si+Si}
and \textit{p+p} collisions at $\sqrt{s_{NN}}$ = 17.3~GeV, as well as central
\textit{Pb+Pb} collisions at $\sqrt{s_{NN}}$ = 12.3, 8.8, 7.6 and 6.3~GeV. Results
of these scans have been compared to UrQMD simulations.\

Our experimental results from two-particle azimuthal correlations show a
flattened away-side peak in central \textit{Pb+Pb} (\textit{Au+Au}) collisions, which
only weakly depends on collision energy over the whole investigated
energy range --- suggesting this effect originates from sources other
than jet-medium interactions. This hypothesis is reinforced by good
agreement with UrQMD regardless of whether jet production was enabled
in the model or not. On the other hand, the amplitude of the near-side
peak in central Pb+Pb (Au+Au) collisions drops visibly with decreasing
collision energy, flattening out around $\sqrt{s_{NN}}$ = 8.8~GeV and turning
into a depletion below that energy --- possibly an effect related to the onset
of deconfinement. Finally, hints of the ridge structure can be observed
in pseudorapidity--azimuth correlation functions from central \textit{Pb+Pb}
collisions at $\sqrt{s_{NN}}$ = 17.3~GeV.
}

\FullConference{5th International Workshop on Critical Point and Onset of Deconfinement - CPOD 2009,\\
		 June 08 - 12 2009\\
		 Brookhaven National Laboratory, Long Island, New York, USA}

\begin{document}

\section{Introduction}
\label{sec:introduction}

Among numerous observables expected to reliably probe the presence of deconfined quarks
and gluons in high-energy heavy-ion collisions, the one most often found ``in the spotlight''
in the past decade has been two-particle azimuthal correlations at high $p_{T}$. For collision energies
lower than those that will become available at the LHC, this approach has always been considered the primary 
way of observing modification of properties of jets --- highly collimated showers of particles,
originating from hard scattering of partons and therefore produced early in a collision --- as a result
of their interaction with the deconfined medium~\cite{WangGyulassy}. Studies along those lines conducted
by RHIC experiments were highly successful: not only did they observe the expected signatures of jet modifications
but also, through further investigations, further evidence was provided which has led to unexpected conclusions
regarding the properties of the observed medium (the ``perfect liquid'' description)~\cite{STARresults, PHENIXresults}.
However, until quite recently (\cite{CERES2005}) studies of this sort were largely neglected
at the CERN SPS.

In early 2008 the NA49 Collaboration presented its first results on two-particle
azimuthal correlations, concluding that the flattening of the away side of the
correlation function in most central \textit{Pb+Pb} collisions at $\sqrt{s_{NN}}$ = 17.3~GeV,
as well as the observed dependence of the near-side amplitude on the charge of trigger
and associate particle, was consistent with qualitative expectations
of QGP presence in central high-energy heavy-ion interactions~\cite{SzubaQM2008}. 

Subsequently, we studied the behaviour of the two-particle azimuthal correlation
function in central collisions on other observables: system size and collision energy.
With deconfinement in heavy-ion collisions believed to set in at low SPS
energies~\cite{NA49onset}, observation of evolution of the correlation function may
shed light on the mechanism responsible for the away-side flattening. For central collisions
subtraction of flow is not necessary, an advantage in view of the criticism of the commonly
employed subtraction techniques~\cite{gyulassyZYAM, trainorZYAM}.

\section{Data Sets and the Analysis Method}
\label{sec:setup_method}

The present analysis is based on the following data sets of NA49: central \textit{Pb+Pb} collisions,
$\sigma/\sigma_{geom}$ = 0--5~\%, at $\sqrt{s_{NN}}$ = 17.3, 12.3, 8.8, 7.6 and 6.3~GeV; central
\textit{Si+Si} collisions, $\sigma/\sigma_{geom}$ = 0--5~\%, at $\sqrt{s_{NN}}$ = 17.3~GeV;
\textit{p+p} collisions ($\approx$90~\% inelastic) at $\sqrt{s_{NN}}$ = 17.3~GeV.

We have compared the experimental correlation functions to results from simulated events using
the string-hadronic model UrQMD~2.3~\cite{urqmd1, urqmd2}, which allows optional incorporation
of jet production from PYTHIA~\cite{pythia}.

\textbf{The method} of calculating two-particle azimuthal correlation functions was described
in detail in our previous report~\cite{SzubaQM2008}. Acceptance-corrected correlation functions $C_{2}(\Delta\phi)$
were obtained in the $\Delta\phi$ range of $\left[ 0,~\pi \right]$. The transverse momentum selections remain
unchanged: $2.5~GeV/c \le p_T^{trg} \le 4.0~GeV/c$ for trigger particles and $1.0~GeV/c \le p_T^{asc} \le 2.5~GeV/c$
for associates.

Additionally, two new techniques have been introduced. The ``central \textit{Pb+Pb} at $\sqrt{s_{NN}}$
= 17.3~GeV'' correlation function used as reference for other functions in the scan was
parametrised with a two-part polynomial fit (third-order on the near side, linear on the away side). This
has greatly improved the ease of comparisons. Secondly, each correlation function
from the energy scan has been fitted with two linear functions (one for the near side, one for the away side),
substituting the comparison of peak values by comparing the slopes of lines fitted to $C_{2}(\Delta\phi)$.

For all azimuthal correlation functions statistical errors are plotted as bars, with systematic uncertainties
illustrated using gray boxes. In case of values extracted from fits, their error bars combine statistical
and systematic uncertainties.

Last but not least, the two-dimensional ($\Delta\eta$, $\Delta\phi$) correlation function is defined in an analogous
way to the azimuthal function, by simply replacing same- and mixed-event $\Delta\phi$ distributions 
used in calculations by their ($\Delta\eta$, $\Delta\phi$) counterparts (see Equation~\ref{eqn:c2etaphi}):

\begin{equation}
  C_{2}(\Delta\eta, \Delta\phi) = \frac{N_{corr}(\Delta\eta, \Delta\phi)}{N_{mix}(\Delta\eta, \Delta\phi)}
    \frac{\int{N_{mix}(\Delta\eta', \Delta\phi')}\mathrm{d}(\Delta\eta')\mathrm{d}\Delta\phi')}
      {\int{N_{corr}(\Delta\eta', \Delta\phi')}\mathrm{d}(\Delta\eta')\mathrm{d}(\Delta\phi')} .
  \label{eqn:c2etaphi}
\end{equation}

\section{Results}
\label{sec:results}

\subsection{System-size Scan}
\label{sec:systemSizeScan}

\begin{figure}[htb]
  \begin{center}
    \includegraphics[width=0.49\textwidth]{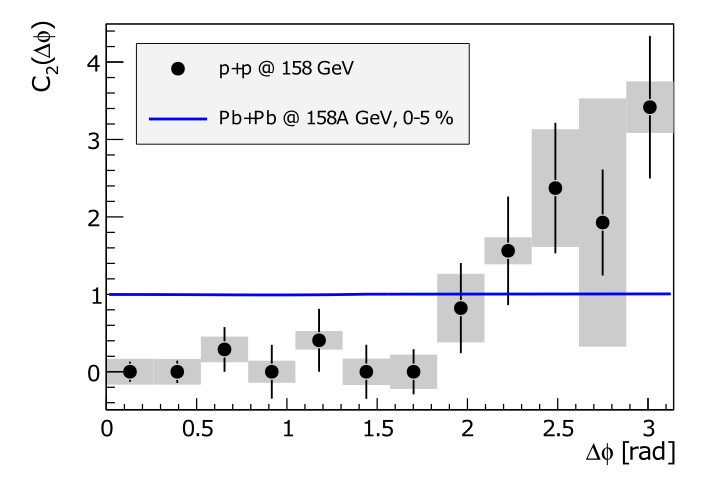}
    \includegraphics[width=0.49\textwidth]{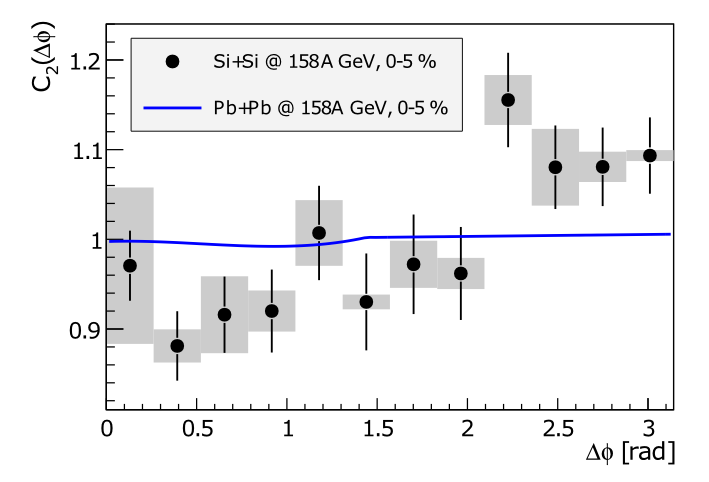}
  \end{center}
  \caption{Two-particle correlation functions from \textit{p+p} (left) and central \textit{Si+Si} (right) events
    at $\sqrt{s_{NN}}$ = 17.3~GeV, compared to a parametrisation of central-\textit{Pb+Pb} results at the same energy (curves).}
  \label{fig:systemSizeScan}
\end{figure}

Figure~\ref{fig:systemSizeScan} shows two-particle azimuthal correlation functions from \textit{p+p}
and central \textit{Si+Si} collisions at $\sqrt{s_{NN}}$ = 17.3~GeV, compared to a parametrisation of
\textit{Pb+Pb} results~\cite{SzubaQM2008}. Overall strength of the correlation becomes significantly larger
as the system size decreases. Moreover, no flattening of the away-side peak, as present in central heavy-ion events,
is visible in \textit{Si+Si} and \textit{p+p} collisions --- indeed, the peak becomes narrower
with decreasing system size.

\subsection{Energy Scan}
\label{sec:energyScan}

\begin{figure}[htb]
  \begin{center}
    \includegraphics[width=0.9\textwidth]{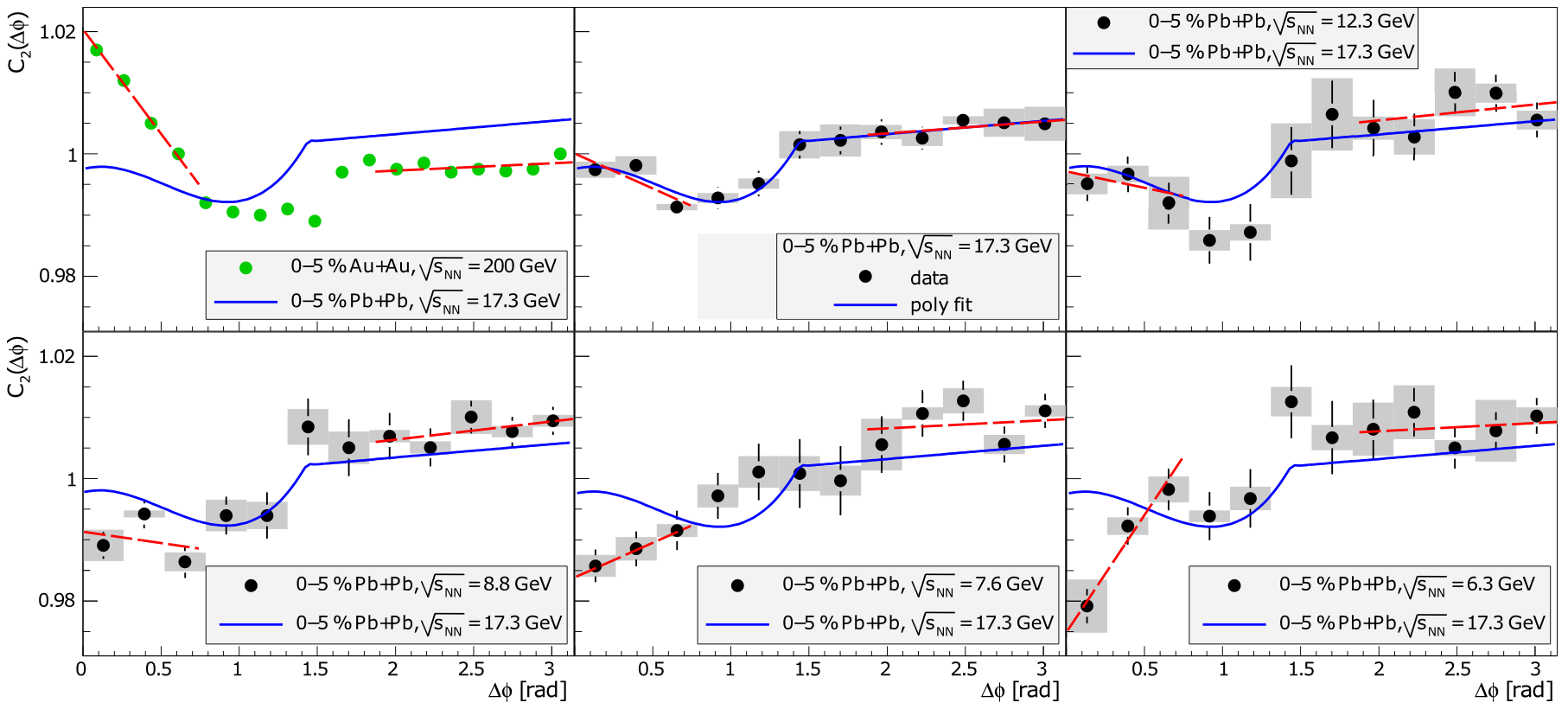}
  \end{center}
  \caption{Two-particle correlation functions from central \textit{Pb+Pb} (\textit{Au+Au}) events at $\sqrt{s_{NN}}$ = 
    200, 12.3, 8.8, 7.6 and 6.3~GeV compared to results from central \textit{Pb+Pb} collisions at 17.3~GeV. Dashed red
    lines illustrate linear fits used to extract slope parameters, plotted in Figure~\protect\ref{fig:urqmd_Escan}.}
  \label{fig:energyScan}
\end{figure}

In Figure~\ref{fig:energyScan} the correlation function from central \textit{Pb+Pb} collisions
at $\sqrt{s_{NN}}$ = 17.3~GeV is compared to results from the same system at 12.3, 8.8, 7.6 and 6.3~GeV;
for illustration we also included a function obtained by PHENIX at the RHIC (\textit{Au+Au} collisions at $\sqrt{s_{NN}}$ = 200~GeV),
for the same centrality and $p_{T}$ ranges~\cite{phenixFcn}. Dashed red lines depict linear fits of the near and away sides
of $C_{2}(\Delta\phi)$. The resulting slope parameters are plotted in Figure~\ref{fig:urqmd_Escan}. The near-side peak appears
to turn into depletion with decreasing energy, whereas shape and amplitude of the away-side enhancement remains mostly unchanged
throughout the scan. Should the flattening of the latter be considered a quark-gluon plasma signature, these results are at odds with present-day
expectations that the QGP is produced only at higher energies.

\subsection{$p_T$ Scan}
\label{sec:pTscan}

\begin{figure}[htb]
  \begin{center}
    \includegraphics[width=0.9\textwidth]{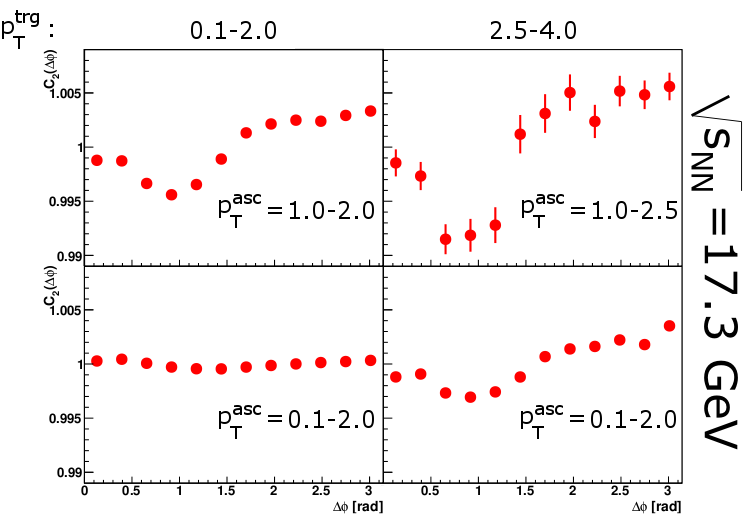}
    \includegraphics[width=0.9\textwidth]{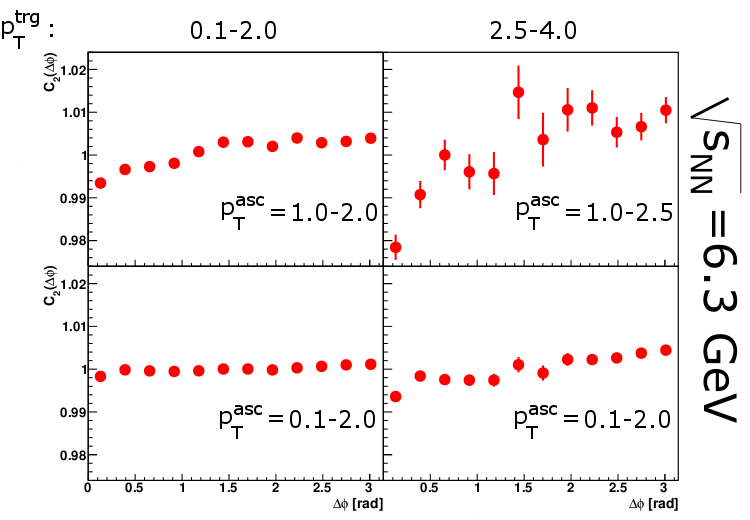}
  \end{center}
  \caption{Two-particle correlation functions from central \textit{Pb+Pb} collisions at $\sqrt{s_{NN}}$ = 17.3~GeV
    (top) and 6.3~GeV (bottom), for different ranges of trigger and associate $p_T$. Only statistical errors are shown.}
  \label{fig:pTscan}
\end{figure}

In light of observed evolution of near-side amplitude, a question arises: can this effect, the peak--depletion transition
near $\sqrt{s_{NN}}$ = 8.8~GeV in particular, be an artifact introduced by the particular selection of transverse-momentum intervals
of the tracks? Therefore, we have performed a comprehensive $p_T$ scan of two-particle azimuthal correlations in central \textit{Pb+Pb}
events at two collision energies from the investigated range: $\sqrt{s_{NN}}$ = 17.3 and 6.3~GeV. A selection of results of this scan
can be found in Figure~\ref{fig:pTscan}, showing the shapes observed at the two energies to be stable over a wide range
of transverse-momentum selection.

\subsection{Comparison with UrQMD}
\label{sec:comparisonWithUrQMD}

\begin{figure}[htb]
  \begin{center}
    \includegraphics[width=0.49\textwidth]{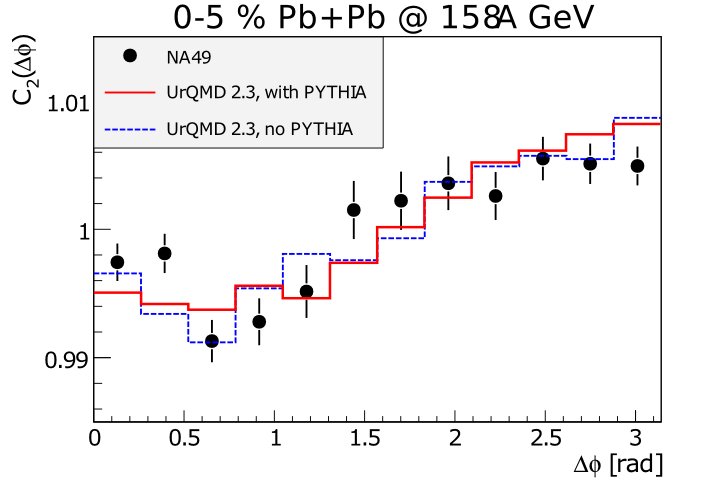}
    \includegraphics[width=0.49\textwidth]{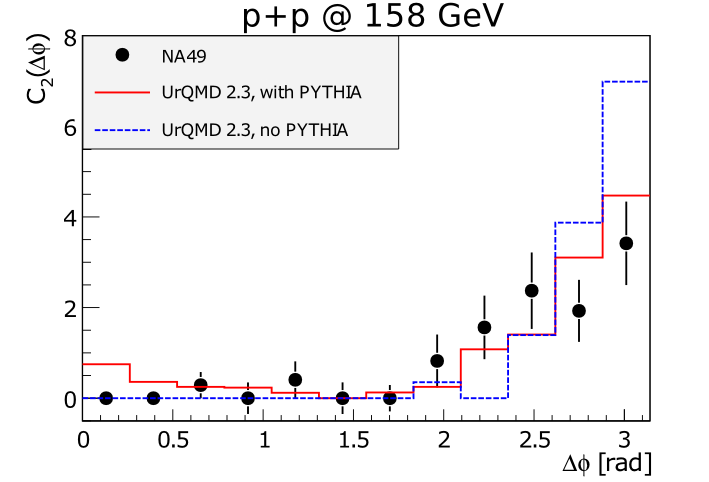}
  \end{center}
  \caption{Comparison of experimental and simulated correlation functions for central \textit{Pb+Pb} (left) and \textit{p+p}
    (right) collisions at $\sqrt{s_{NN}}$ = 17.3~GeV. Black points: experimental data, solid red lines: UrQMD with PYTHIA,
    dashed blue lines: UrQMD without PYTHIA. Systematic errors have been omitted for clarity.}
  \label{fig:urqmd_Pb_p}
\end{figure}

A comparison of azimuthal correlation functions from central-\textit{Pb+Pb} and \textit{p+p} collisions
at $\sqrt{s_{NN}}$ = 17.3~GeV between data and UrQMD simulations at SPS energies can be found in Figure~\ref{fig:urqmd_Pb_p}.
For both systems good agreement can be observed between the data and the simulations, especially on the away side.
Moreover, strong similarity of model correlation functions with and without jet contribution implies this particular
correlation source not to play a major role in the SPS energy range.

\begin{figure}[htb]
  \begin{center}
    \includegraphics[width=0.9\textwidth]{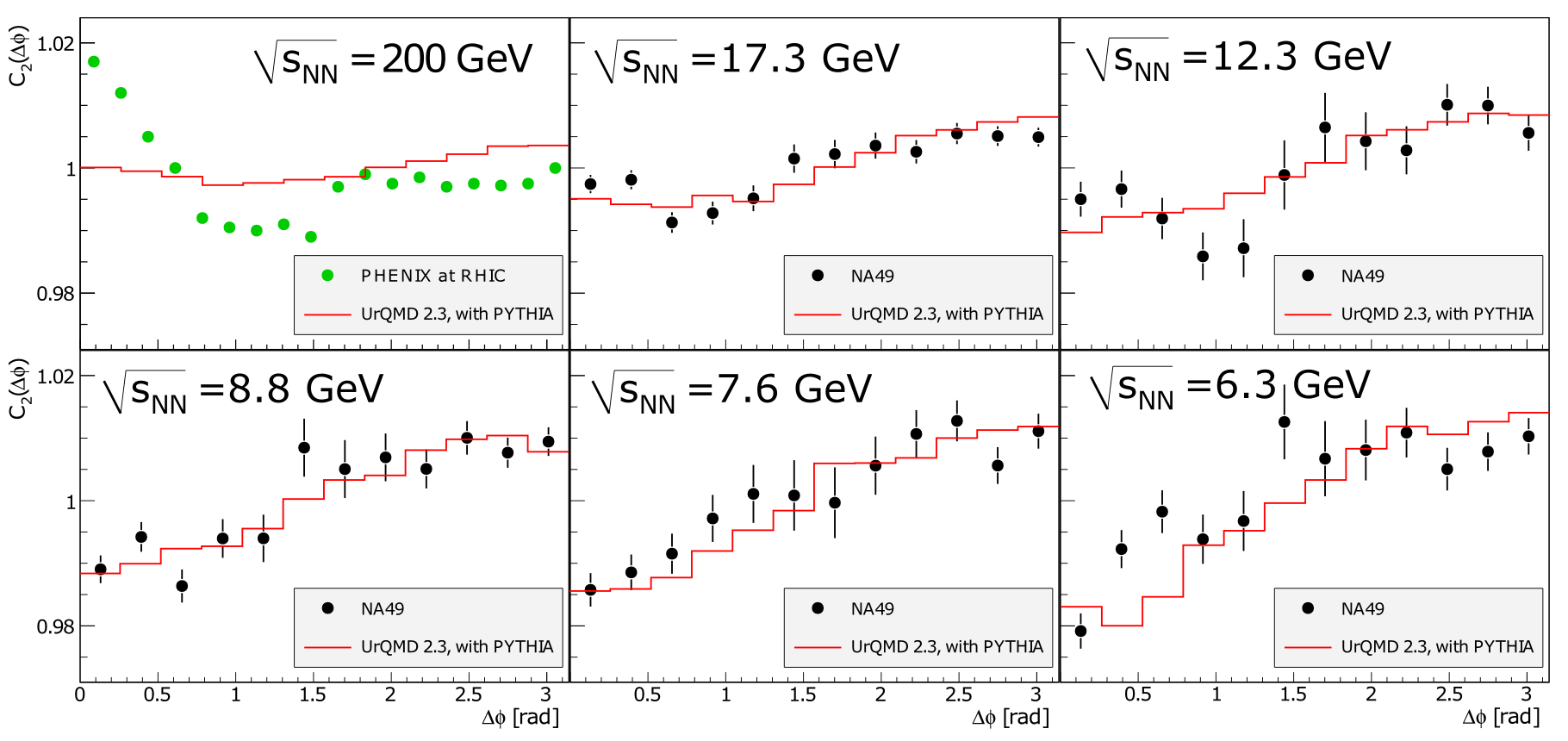}
  \end{center}
  \caption{Experimental two-particle correlation functions from central \textit{Pb+Pb} (\textit{Au+Au}) events
    at $\sqrt{s_{NN}}$ = 200, 17.3, 12.3, 8.8, 7.6 and 6.3~GeV (black points), compared to results from UrQMD (red lines).}
  \label{fig:urqmd_Escan_raw}
\end{figure}

\begin{figure}[htb]
  \begin{center}
    \includegraphics[width=0.49\textwidth]{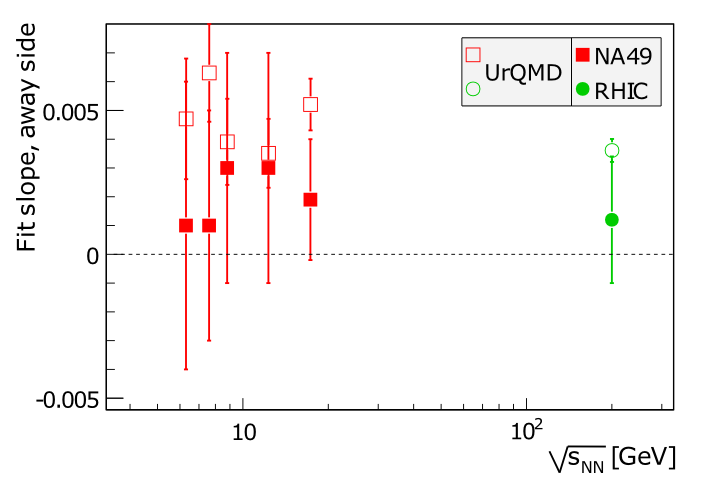}
    \includegraphics[width=0.49\textwidth]{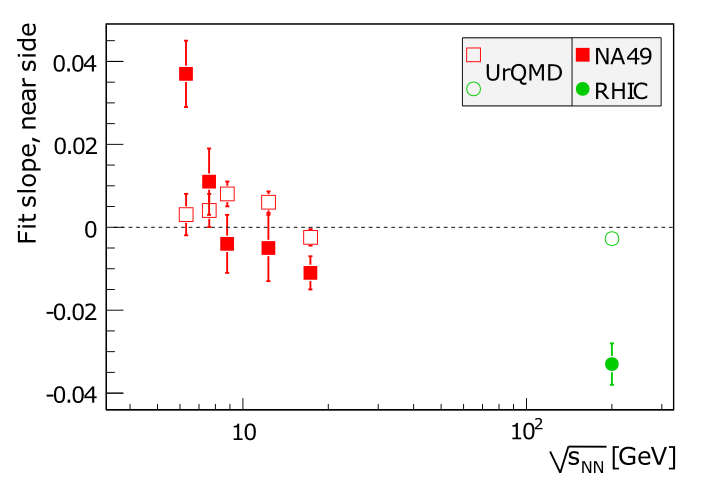}
  \end{center}
  \caption{Dependence of slope values extracted from real-data (full points) and UrQMD (open points) correlation
    functions as a function of collision energy. Left: away side, right: near side.}
  \label{fig:urqmd_Escan}
\end{figure}

Correlation functions for UrQMD data sets at other energies were also produced and can be seen, compared to experimental results,
in Figure~\ref{fig:urqmd_Escan_raw}. Moreover, slopes of linear fits to both UrQMD simulations and real data are shown
in Figure~\ref{fig:urqmd_Escan}. It can clearly be seen here that the weak dependence of away-side slope on energy in real data
is also present in UrQMD simulations. However, data and simulations follow different trends on the near side.

\subsection{($\Delta\eta$, $\Delta\phi$) Correlations}
\label{sec:detadphi}

\begin{figure}[htb]
  \begin{center}
    \includegraphics[width=0.9\textwidth]{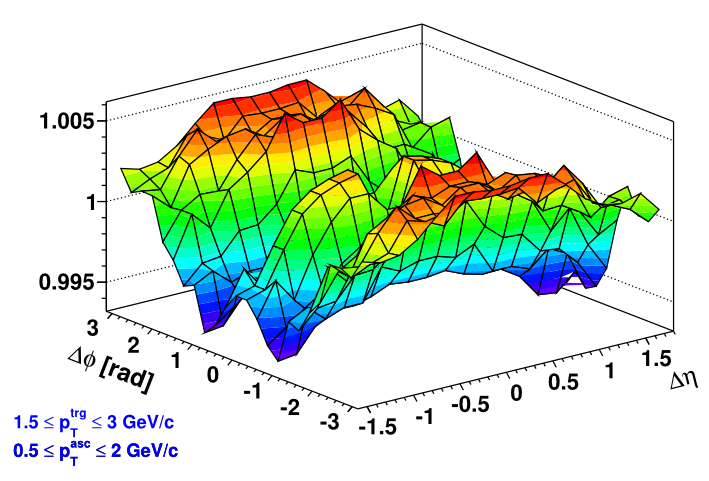}
  \end{center}
  \caption{A two-particle ($\Delta\eta$, $\Delta\phi$) correlation function from central (0--5~\%) \textit{Pb+Pb}
    collisions at $\sqrt{s_{NN}}$ = 17.3~GeV, for $1.5~GeV/c \le p_T^{trg} \le 3.0~GeV/c$
    and $0.5~GeV/c \le p_T^{asc} \le 2.0~GeV/c$. The function was calculated for the (+,+) quadrant, then
    reflected along $\Delta\eta$ = 0 and $\Delta\phi$ = 0.}
  \label{fig:detadphi}
\end{figure}

As mentioned in the introduction, one of the phenomena observed in angular correlations by RHIC experiments
is the so-called ridge: ($\Delta\eta$, $\Delta\phi$) correlation functions from central heavy-ion collisions
at the RHIC show a pronounced enhancement, Gaussian-like in $\Delta\phi$ but uniform in $\Delta\eta$, which is not
visible in nucleon--nucleon or nucleon--nucleus collisions~\cite{starRidge1, starRidge2}. In an attempt to determine
whether a similar structure can be observed in heavy-ion collisions at the SPS, we have produced two-particle
($\Delta\eta$, $\Delta\phi$) correlation functions from central \textit{Pb+Pb} events at $\sqrt{s_{NN}}$ = 17.3~GeV;
one of these functions can be seen in Figure~\ref{fig:detadphi}. Note that in order to reduce statistical uncertainties,
the function in question was calculated only for positive values of both $\Delta\phi$ and $\Delta\eta$; negative quadrants
of the figure are reflections. Moreover, $p_{T}$ ranges used to produce this function were somewhat lower than
those for azimuthal functions, again to reduce uncertainties.

Our results tentatively suggest a ridge-like structure may indeed be present also at high SPS energies. Making a more
definite statement on this subject is unfortunately difficult due to our limited $\Delta\eta$ range, the fact the near-side
peak is much wider in this direction than at the RHIC and, last but by not least, the limited event statistics.

On the other hand, a dip-like structure is very clearly visible around the (0,0) point of the correlation function.
At this time the origin of this dip is not known.

\section{Summary}
\label{sec:summary}

A system-size and energy scan of two-particle correlation functions at high $p_{T}$ was performed by NA49.
A flattening of the function's away side was observed in central heavy-ion collisions even
at low SPS energies, raising doubts about the standard parton energy loss interpretation. Interestingly,
UrQMD predictions agree well with the away-side experimental results.

On the other hand, clear energy dependence of the correlation function was observed
on the near side. As the energy decreases an enhancement changes into a depletion in the region
close to the probable onset of deconfinement; this effect appears to be independent of trigger and associate
$p_{T}$ ranges and is not present in UrQMD model calculations. Whether the observed behaviour is indeed related
to the onset of deconfinement or simply coincidental remains to be determined, through further studies.

Last but not least, our ($\Delta\eta$, $\Delta\phi$) results tentatively suggest the existence
of the ridge in central heavy-ion collisions at the top SPS energy. In light of our observations from azimuthal
correlations, this structure --- if it indeed exists --- would not be related to quark-gluon plasma; this is
consistent with present-day belief that the ridge is much more complicated than originally thought~\cite{moschelliRidge, shuryakRidge}.
Moreover, a dip can be observed near the (0,0) point of the correlation function, whose origin is yet to be determined.


\begin{thebibliography}{99}
  \bibitem{WangGyulassy} X.-N.~Wang, M.~Gyulassy, Phys.~Rev.~Lett.~\textbf{68}, (1992) 1480-1483.
  \bibitem{STARresults} J.~Adams \textit{et~al.} (STAR Collaboration), Phys.~Rev.~Lett.~\textbf{95}, (2005) 152301.
  \bibitem{PHENIXresults} C.~Zhang for the PHENIX Collaboration, J.~Phys.~G \textbf{34}, (2007) S671-S674.
  \bibitem{CERES2005} M.~Ploskon for the CERES Collaboration, Acta~Phys.~Hung.~\textbf{A27}, (2006) 255-258.
  \bibitem{SzubaQM2008} M.~Szuba for the NA49 Collaboration, arXiv:0805.4637 [nucl-ex], to be published in Indian Journal of Physics.
  \bibitem{NA49onset} C.~Alt \textit{et~al.} (NA49 Collaboration), Phys.~Rev.~\textbf{C77}, (2008) 024903.
  \bibitem{gyulassyZYAM} M.~Gyulassy, talk at the final workshop of the Virtual Institute of Strongly Interacting Matter,
    University of Frankfurt, 19 May 2008
  \bibitem{trainorZYAM} T.A.~Trainor, arXiv:0904.1733 [hep-ph]
  \bibitem{urqmd1} S.A.~Bass, M.~Belkacem, M.~Bleicher, M.~Brandstetter, L.~Bravina, C.~Ernst, L.~Gerland, M.~Hofmann,
    S.~Hofmann, J.~Konopka, G.~Mao, L.~Neise, S.~Soff, C.~Spieles, H.~Weber, L.A.~Winckelmann, H.~Stocker, W.~Greiner,
    Ch.~Hartnack, J.~Aichelin and N.~Amelin, Prog. Part. Nucl. Phys. 41, (1998) 225-370.
  \bibitem{urqmd2} M.~Bleicher, E.~Zabrodin, C.~Spieles, S.A.~Bass, C.~Ernst, S.~Soff, L.~Bravina, M.~Belkacem,
    H.~Weber, H.~Stocker, W.~Greiner, J. Phys. G: Nucl. Part. Phys. 25, (1999) 1859-1896.
  \bibitem{pythia} T.~Sj\"{o}strand, S.~Mrenna, P.~Skands, JHEP 0605, (2006) 026.
  \bibitem{phenixFcn} S.S.~Adler \textit{et~al.} (PHENIX Collaboration), Phys.~Rev.~Lett.~\textbf{97}, (2006) 052301.
  \bibitem{starRidge1} J.~Adams \textit{et~al.} (STAR Collaboration), Phys.~Rev.~Lett.~\textbf{95}, (2005) 152301.
  \bibitem{starRidge2} J.~Putschke, J.~Phys.~\textbf{G34}, (2007) S679-S683.
  \bibitem{moschelliRidge} G.~Moschelli, S.~Gavin, talk at the RHIC \& AGS Users' Meeting, June 2009
  \bibitem{shuryakRidge} E.~Shuryak, these proceedings
\end{thebibliography}
\end{document}